\documentclass[12pt]{article}
\usepackage[english]{babel}

\usepackage[letterpaper,top=4cm,bottom=4cm,left=3cm,right=3cm,marginparwidth=1.75cm]{geometry}
\usepackage{setspace}
\usepackage{amsmath}
\usepackage{graphicx}
\usepackage[colorlinks=true, allcolors=blue]{hyperref}
\usepackage{natbib}

\title{Safeguarding Autonomy: a Focus on Machine Learning Decision Systems}
\author{
    Paula Subías-Beltrán\textsuperscript{1,2} \and
    Oriol Pujol\textsuperscript{3} \and
    Itziar de Lecuona\textsuperscript{2,4}
}
\date{}

\begin{document}
\maketitle
\footnotetext[1]{Eurecat, Centre Tecnològic de Catalunya, Barcelona, Spain.}
\footnotetext[2]{Bioethics and Law Observatory - UNESCO Chair in Bioethics, Universitat de Barcelona, Barcelona, Spain.}
\footnotetext[3]{Dept. de Matemàtiques i Informàtica, Universitat de Barcelona, Barcelona, Spain.}
\footnotetext[4]{Dept. of Medicine, Universitat de Barcelona, Barcelona, Spain.}
\footnotetext[5]{Corresponding author: Paula Subías-Beltrán, \texttt{paulasubias@ub.edu}.}

\begin{abstract}
As global discourse on AI regulation gains momentum, this paper focuses on delineating the impact of ML on autonomy and fostering awareness. Respect for autonomy is a basic principle in bioethics that establishes persons as decision-makers. While the concept of autonomy in the context of ML appears in several European normative publications, it remains a theoretical concept that has yet to be widely accepted in ML practice. Our contribution is to bridge the theoretical and practical gap by encouraging the practical application of autonomy in decision-making within ML practice by identifying the conditioning factors that currently prevent it. Consequently, we focus on the different stages of the ML pipeline to identify the potential effects on ML end-users' autonomy. To improve its practical utility, we propose a related question for each detected impact, offering guidance for identifying possible focus points to respect ML end-users autonomy in decision-making.
\end{abstract}

\textbf{keywords}: machine learning, human autonomy, AI ethics, bioethics and human rights.

\section{Introduction}
Autonomy alludes to the ability to make decisions on one's own while respecting the authenticity of one's own motives.
In the literature, it tends to be associated with self-determination and self-respect, whose execution comes at the sacrifice of devoting time to gathering all the knowledge required to behave in accordance with each own's beliefs and ideals. Full autonomy entails total control over all facets of our existence, which can be a time-consuming process and unrealizable in practice.

Autonomy is essential to the development of individuals, societies, and democracies, but in order to keep progressing we need to find a balance between taking action and controlling our surroundings. There is tension between giving up our agency and being active agents in pursuing our goals. We can give up part of our autonomy by taking advantage of existing services and technologies that allow us to perform certain actions without our active participation. But, our goal is to be able to decide when we give up our autonomy and in what form.

Everyday systems based on Machine Learning (ML) technology are gaining more and more space in our lives and, in turn, increasing their effect on society. Now, what role do we intend for this technology to fulfill?
After several years of normative overproduction, we are now at a point where the regulation of Artificial Intelligence (AI) is about to become a reality. This step reflects the growing awareness of the need to frame ML development within a shared ethical framework. The ethical framework that is socially acceptable and preferable will be determined by the experience of end-users, who will ultimately shape the ethical standards that are deemed acceptable within their social context \citep{Floridi2018SoftRegulation}. Thus, to achieve an actionable regulation, it is necessary to outline how to act on each of the ethical principles that we consider relevant to the progress of society.

While the concept of respecting the autonomy of end-users in the context of ML appears in a variety of normative publications~\citep{UNESCO2022RecommendationIntelligence,EuropeanCommission2021ProposalActs}, it is still a theoretical concept that has yet to be broadly accepted in the ML practice, as evidenced by the challenges one encounters when attempting to contest the results generated by ML models. This manifests a systematical disdain for our autonomy, undermining the legitimacy of the contemporary debate on the subject. 
Incorporating ethics from the start helps avoid the weakness of the present discourse, which promotes the adoption of safeguards against ML systems that may impair person's autonomy, while simultaneously recognizing the autonomous behavior of such ML systems~\citep{CommitteeonArtificialIntelligence2023DraftLaw}.

This work seeks to aid a wide range of audiences, from ML practitioners to ML auditors, 
by giving guidelines for uncovering the impact of ML on users' autonomy.
The process of identifying factors to watch is deliberately organized under a simplified framework that depicts key phases of the ML pipeline. This proposed structure aims to improve understanding among all of the parties engaging in the ML ecosystem. Thus, we hope to bridge the gap between theoretical discussions regarding autonomy and the practical considerations required in its implementation. It intends to make the identification of points of attention more accessible and useful, establishing a shared understanding that will benefit both those engaged in philosophical discourse and those actively involved in the practical implementation of ML technologies. In essence, one of the tangible outcomes of this paper is a series of questions that can be used to assess the degree of respect for autonomy in the lifecycle of ML systems.

With the ultimate goal of building a diagnostic framework, we conduct an analysis to identify key points of the exercise of autonomy through decision-making that may emerge across the ML pipeline. This endeavor is rooted in a comprehensive examination of philosophical perspectives on autonomy and a thorough understanding of ML practices. This approach not only strengthens the diagnostic but also ensures that the framework is founded on a good understanding of the fundamental notion of autonomy. Furthermore, the series of questions proposed can serve as an effective starting point for raising awareness about the impact of ML on autonomy. This will help to promote transparency and proactive reporting, which are essential elements of the AI Act.
This manuscript is structured in four additional sections. For the sake of clarity, we will use the term AI to encompass the overarching technology, while ML will precisely designate the currently employed solutions. $\S$\ref{sec:autonomy} discusses the concept of autonomy and examines several philosophical approaches that describe its main components. The goal of this section is twofold: to provide insight into the complex nature of autonomy while also shedding light on the similarities that exist across various philosophical views. $\S$\ref{sec:risks} examines the key considerations involved in examining the impact of ML on end-user autonomy. Its goal is to create the groundwork for the diagnostic procedure presented in the $\S$\ref{sec:autonomy-in-ML}. This section's analysis goes over the main steps of the ML pipeline, identifying important implications for autonomy and presenting a diagnostic question for further investigation. Finally, the manuscript concludes with a discussion and summarizing remarks in $\S$\ref{sec:conclusions}.

\section{Autonomy and decision-making in ML}
\label{sec:autonomy}
The concept of autonomy is often used in European regulations, yet there is no agreement on its precise definition. In essence, all theories recognize the capacity for self-determination as a fundamental aspect of autonomy.

Every person has a self-\textit{authority} that is derived from their innate capacity to take the initiative and take action on their own. A person must intrinsically acknowledge their judgment as authoritative before they may form an intention to do one thing and not another, regardless of whether they should only follow their own judgment or someone else's~\citep{Buss2018PersonalAutonomy}. 
Despite the inalienable nature of our authority over ourselves, it is possible to fail to govern ourselves. Authority is a form of self-government, but it is no guarantee that whenever we act, our actions are propelled by the influence of our power to decide what course of action to pursue.
Individuals who fail to exercise self-governance in their behaviors demonstrate a lack of power to direct their actions~\citep{Fischer1982ResponsibilityControl}. Yet, given that humans have inherent authority over themselves, such a scenario is unattainable. However, it is critical to acknowledge that people might succumb to cravings or urges that, at times, overpower more fundamental beliefs and influence their actions.

This is related to \cite{Kant1785FundamentalMorals}'s conceptualization of the will in his well-known manuscript ``The Groundwork of the Metaphysics of Morals''. In his opinion, having a will equates to being rational, and a free will is one that is unaffected by external factors. According to him, the will is guided by both reason and inclination. As long as our behaviors are consistent with our own imperatives, we have control over our moral law.
\cite{Noggle2022TheManipulation} offers a similar perspective on the impact of manipulation on autonomy. He contends that it is a common idea that manipulation is morally unacceptable because it subverts autonomous choices. Nonetheless, he identifies cases in which manipulation may actually improve autonomy. Some authors go so far as to claim that manipulation does not deprive victims of their authority, refuting the notion that autonomous individuals would logically refuse exposure to manipulative effects~\citep{Buss1992ManipulationContext, Buss1987TacticsManipulation}.

One important aspect of autonomy is the idea of \textit{authenticity}, which is most commonly understood as owning up to what one is and does~\citep{Heidegger1962BeingTranslation}. Heidegger interprets people as ``relation of beings'', a relation established by one's present self and the potential one can and will become, unfolding over time within an open realm of possibilities. This ongoing construction of ourselves is the direct effect of continually taking a stance on who we are. People can constantly act according to one's overarching life project~\citep{Guignon2004OnAuthentic} or act as one of the herd, be adrift, falling into what Heidegger refers to as a mode of existence. This distinction shows different modes of living: an authentic way of life is owned, an average way of life is unowned, and an inauthentic way of life is disowned~\citep{Varga2023Authenticity}.

\textit{Agency} is another essential aspect of autonomy. It represents the underlying intention of an action. Its analysis is two-fold~\citep{Schlosser2019Agency}: the standard conception construes actions in terms of intentionality and the standard theory of action explains the intentionality of action in terms of causation by the agent's mental states and events. The former is frequently seen in two ways: either as intentions dependent on the explanation given or as perceiving an action's intention as adhering to a reasonable practical syllogism, typically in line with the person's goals~\citep{Anscombe1957Intention,Davidson1963ActionsCauses}. The standard theory of action, also known as the causal theory of action, often provides a means-end rationale to explain the motive for an action. Nonetheless, this has been criticized for the possible difficulty agents may face in identifying with a particular motive~\citep{Velleman1992WhatActs}. This view is referred to as procedural independence because it accounts for the procedure by which people come to identify a desire as their own~\citep{Dworkin1976AutonomyControl}.

Autonomy is an interdependent value intricately connected to the human rights framework \citep{Wright2014SurveillanceEurope}, with values such as privacy, justice, and transparency. A loss of privacy is frequently linked to a decline in autonomy, e.g., protecting people's privacy secures the creation of their personal identity, which defines each person's beliefs and desires~\citep{Floridi2006FourPrivacy, Floridi2005ThePrivacy}. Additionally, autonomy is closely intertwined with justice, which dictates that individuals should not face unfair discrimination. When individuals are empowered to make their own decisions, it fosters a sense of justice and equality, as everyone is given the same opportunity to exercise their rights and duties. In this sense, when we treat persons equally, we guarantee that everyone has the opportunity to exercise their capacity to decide for themselves how to live their lives and to revise it as needed \citep{Richards1971AAction}. Furthermore, transparency is a vital element in the nurturing of autonomy, as it serves to establish a foundation of trust and facilitate informed decision-making. When individuals and groups have clear access to information, they are able to comprehend the context and rationale behind decisions, thereby empowering them to act independently and with confidence.
By recognizing that autonomy cannot exist in isolation, the importance of a supportive societal framework is emphasized. This interconnectedness means that protecting an individual's autonomy also involves safeguarding their privacy, while ensuring justice and transparency, among others. Thus, a truly autonomous society is one where these values are upheld collectively, fostering an environment where all individuals can freely exercise their autonomy.

\subsection{On decision-making}
Autonomy is a multifaceted concept, but in the world of decision-making, it is primarily manifested by the three essential notions mentioned above: authority, authenticity, and agency. Each of these factors makes an important contribution to the decision-making process in the context of ML. However, our primary focus in this work will be on agency, as it is inextricably linked to the purposeful nature of decisions. The ability to act with agency indicates a sense of purpose and direction in decision-making within ML methods.

Nonetheless, it is critical to recognize the roles of authority and authenticity in this setting. Authority plays an important role in defining the decision-making environment by indicating the source or sources of influence behind decisions. Authenticity is also important in ensuring that decisions are consistent with genuine ideals and principles, which contributes to the integrity of the decision-making process.

Whenever we make decisions, we exercise our autonomy. The ``four-abilities model'' \citep{Grisso1998TheProfessionals} is a well-known model that aims at characterizing the decision-making capacity. It includes choice (the ability to express or communicate one's decision), understanding (comprehension and knowledge or cognition of facts), appreciation (awareness of the nature and significance of the decision), and reasoning (keeping consistency and the ability to draw inferences from premises). Other authors have proposed the inclusion of further abilities, like subjects' emotions, values, and authenticity. However, this paradigm is sufficient for mapping the various decision-making capacities in the exercise of autonomy.

Based on prior philosophical considerations of autonomy and decision-making, we can now identify different stages in its practical application. These can be divided into two stages: first, the development of the skills required to be recognized as an autonomous individual, and second, the execution of decisions, which can be further dissected into three blocks: choice is about \textit{making a decision}, understanding and appreciation can be mapped to \textit{being well-informed}, and reasoning refers to \textit{forming the decision space}.
The goal of this mapping is to demonstrate that the two models under consideration may be effortlessly integrated and are equally relevant to the point we want to communicate. Our preference goes toward the latter model with three different blocks, owing to its simplicity and use of concepts that are more closely related to the practical realization of autonomy.

The chosen model, introduced in~\citep{Subias-Beltran2022TheLearning, Subias-Beltran2023RespectPipeline}, is enhanced with an additional stage focused on developing the skills necessary for individuals to achieve recognition as autonomous. \textit{Competency building and interpersonal duties} are essential components of becoming autonomous individuals~\citep{Laitinen2021AIAutonomy}. Competency development entails learning and improving the skills and abilities required for autonomy, while interpersonal duties refer to an individual’s responsibilities and obligations to others. \textit{Being well-informed} stands for having adequate information to create a diversity of options and evaluate their implications. This helps people to consider a variety of choices and possibilities before making a decision. To make a choice, you must have a collection of prospective possibilities from which to choose. \textit{Forming the decision space} entails identifying all of the potential options for a given situation. To exercise decision-making abilities and freely act, an individual must have a variety of options to choose from as well as the ability to evaluate those options based on their own preferences and values. Individuals must also be free of external pressures that might impair their decision-making abilities or prevent them from \textit{making their decisions}, such as compulsion or manipulation. 
This model will serve as the study's decision-making framework to analyze the effects of ML on autonomy.

\section{How can ML impact autonomy on decision-making?}
\label{sec:risks}
ML possesses the capacity to impact people's autonomy by molding their decision-making procedures. In the framework of this paper, we have identified several areas of vigilance that are essential to comprehending and mitigating possible issues related to the influence of ML on personal autonomy.

\paragraph{Fail to develop competence of autonomy} 
People's actions and decisions are influenced by prevailing norms, values, and societal expectations, revealing the complex interplay between individuals and the larger cultural environment. This dynamic interaction highlights how cultural variables greatly contribute to shaping and steering different facets of human behavior. 
ML has become a substantial technology for our society, exerting a dynamic influence on human behavior. It has the capacity to influence people's perceptions of themselves, their access to knowledge, and the importance they place on societal values. The broad use of these solutions raises the question of whether users of ML tools will reach their full potential, especially in a world where a growing number of tools are constantly streamlining and making their environment simpler. For example, ML is expected to have a stronger impact on children's play, education, information consumption, and general knowledge. This underlines the importance of interested parties working together to analyze the contingencies of employing such technologies, as well as exploring opportunities to apply ML in ways that benefit children's welfare through an intentional and structured approach~\citep{UNICEFInnovation2019MemorandumRights}.

\paragraph{Over-reliance}
The promises of AI refer to the high expectations and possible benefits associated with AI technologies. However, the notion that ML systems have the potential to solve all problems can be misleading. The concept of operating these systems ``as if ML knew it all'' implies a dependence on technology that may not always coincide with its actual capabilities, underlining the need for a nuanced understanding of ML's limitations to ensure reasonable expectations and optimal use.
Excessive reliance implies the cession of autonomy to act by default. There are two ways that an excessive cession of autonomy can happen: forced over-reliance and passive over-reliance. Forced over-reliance is when individuals are compelled to rely excessively on a system, e.g., individuals need to use a solution to keep being part of a system and not suffer social exclusion. On the other hand, passive over-reliance occurs when people depend too heavily on a system because they are overconfident or have technology apathy. In this situation, people might place too much assurance in the system and refrain from acting independently, either genuinely or out of weariness from micromanaging. This has already been shown to be a problem by Robinette et al., who executed an experiment where humans over-trusted robots in situations of emergency evacuation~\citep{Robinette2016OvertrustScenarios}.

\paragraph{Heteronomy dynamics}
The prevalence of ML system's outcomes driving decisions rather than personal choices corresponds to heteronomy dynamics. In this context, heteronomy dynamics refers to the interplay between humans and ML where ML \textit{makes} decisions on behalf of the user, resulting in a loss of user's autonomy. In this scenario, ML end-users perception is influenced by what the ML system delivers, indicating a possible shift in autonomy from human decision-making to blind reliance on the ML system's outcomes. This raises concerns regarding the extent to which individuals should relinquish decision-making authority to ML, as well as the ramifications for personal autonomy in decision-making processes.
The current prevalent lack of transparency in ML solutions amplifies the influence of this heteronomy. As a result, the matter extends beyond concerns about over-reliance on ML to include how ML influences the formulation and rationale for decision-making. For example, low representativeness in the input data limits the scope of the information available to the ML system, hence limiting the resulting outcomes provided to the end user. Because of this, the perspective on how ML is developed and what it processes can reinforce an ML tunnel vision, enabling the end user to believe that the ML's outcomes encompass all available options.

\paragraph{Misalignment} 
According to the dominant narrative, AI systems should not have intrinsic values; this idea is sometimes referred to as the ``alignment problem''. In contrast, we suggest that there is a mismatch between the data these systems employ and the real condition of the world as well as our idealized version of it. This implies that the task at hand goes beyond simply integrating ML with preset values; rather, it entails navigating the intricate differences between the world we live in, our idealized objectives, and the data given into these systems. Achieving significant and useful ML applications requires addressing this subtle imbalance. This is exemplified in the case of criminal justice predictions. It is common to aim to forecast recidivism itself, but the available data do not support this. Instead, the data capture rearrest and reconviction, which is fundamentally different since this data is gathered as a byproduct of police activity, meaning that it only represents instances of crime that are known to the police~\citep{Lum2016ToServe}.

In the next section, we will undertake a thorough evaluation of the significant issues that have been identified, progressing through the steps of the ML pipeline. This extensive analysis will not only shed light on the nature of these major aspects to monitor but will also provide particular examples and questions to help uncover them. By focusing on each stage of the ML pipeline, we hope to provide a more nuanced view of how these contingencies can materialize.

\section{Dissecting the effect on autonomy in the ML pipeline}
\label{sec:autonomy-in-ML}
\begin{figure}[h]
    \centering
    \includegraphics[width=0.75\textwidth]{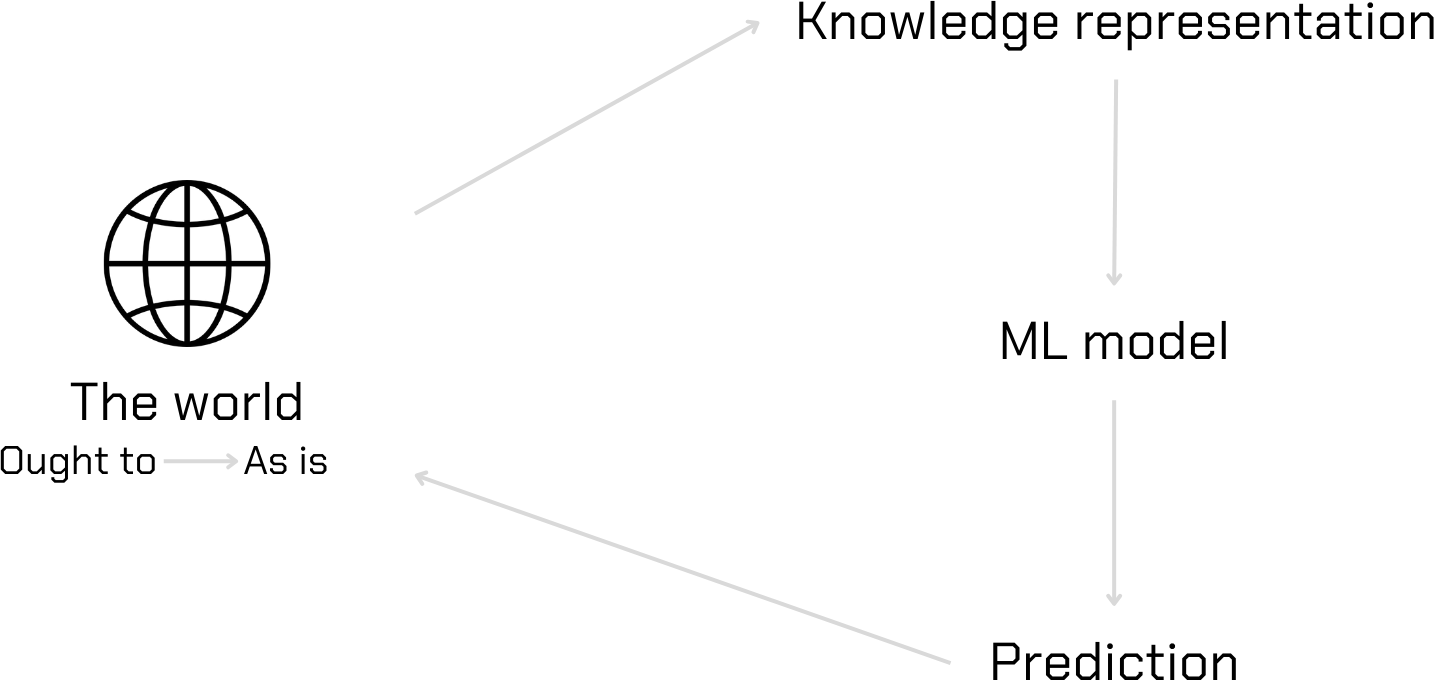}
    \caption{A simplified version of the ML pipeline.}
    \label{fig:ML pipeline}
\end{figure}

In this section, we examine the influence on autonomy as we progress through the various stages of the ML pipeline. There is no single standard for defining the ML pipeline, but given the broad scope of this publication, we choose to compress ML phases in order to have an aggregate discourse, acknowledging there may be different ways to do so, but this facilitates the most relevant points for decision-making and its impact on autonomy. We propose a simplified ML pipeline, as depicted in Figure~\ref{fig:ML pipeline}.

It consists of four phases:
\begin{itemize}
    \item \textbf{The world.} At this stage the main ML task is scoping. Reality \textit{as is} demarcates what exists, shaping and constraining future possibilities. However, this phase is also the moment to contemplate what \textit{ought} to be.
    
    \item \textbf{Knowledge representation.} Reality is quantified through data, and it entails processes such as experimental design and sampling, measurement and digitalization, knowledge representation, data quality, preprocessing and data curation, and labeling if required.
    
    \item \textbf{ML model.} This phase includes all processes related to the modeling, the learning process, and all their dependencies, including metrics and the usage of third-party components.
    
    \item \textbf{Prediction.} This is the application of the ML model and its impact on the world. It includes the process of deploying trained ML models into real-world applications to perform the designed task on new data, the measurement of its impact, and the performative effect on the world.
\end{itemize}

\begin{figure}[h]
    \centering
    \includegraphics[width=\textwidth]{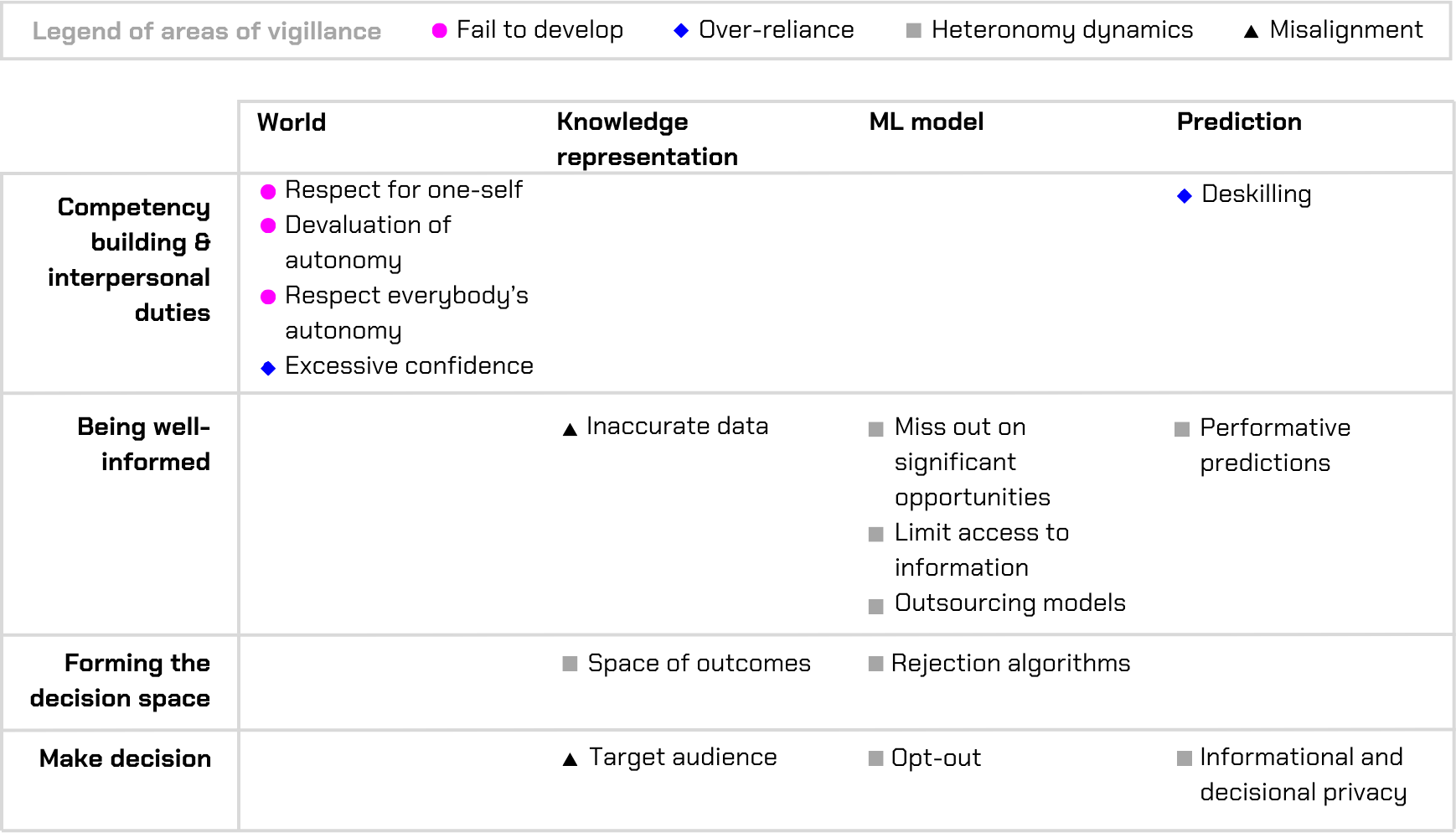}
    \caption{Framework for investigating the impact on autonomy via the lens of the ML pipeline. The area of vigilance is indicated based on color and marker shape.}
    \label{fig:autonomy-ML}
\end{figure}

Crossing these four steps of the ML pipeline with the aforesaid autonomy and decision-making framework ($\S$\ref{sec:autonomy}), allows for the creation of Fig.~\ref{fig:autonomy-ML}. This illustration serves as a reference framework for investigating the impact on autonomy via the lens of the ML pipeline.

We begin by examining potential effects to autonomy and proposing a guiding question to direct the audit of ML solutions on their impact on autonomy. While this work is designed to benefit a diverse range of stakeholders, the questions posed should be directed to ML practitioners, as they are the ones who ultimately implement the design decisions made. The questions proposed do not aim at covering all the details of major areas of vigilance identified in $\S$\ref{sec:risks}, but as a proxy to evaluate how they can materialize in practice.

\subsection{Impact on the world}
ML has the potential to obstruct people's efforts to cultivate self-determination in a number of ways. This obstacle is especially significant at this particular step of the ML pipeline since it is during this stage that people develop the foundational abilities needed for autonomy. This includes impediments that not only inhibit the cultivation of self-determination, but also change the dynamics of how humans interact with ML solutions. Exploring these factors is critical for gaining a thorough knowledge of the problems presented at this part of the pipeline. The following dimensions to evaluate are identified in this phase of ML.

\begin{itemize}
    \item \textbf{Respect for one-self.} ML solutions have considerably accelerated the flow of information available to individuals, as seen by their strong use in recommendation systems. These systems use ML algorithms to assess user preferences and habits resulting in individualized suggestions and content. As a result, consumers have more streamlined and customized access to a wealth of information. This wealth of information has resulted in a poverty of attention \citep{Simon1971DesigningWorld,Goldhaber1997TheNet}. Because we are unable to generate opinions on every occurrence, we outsource this responsibility through interpassivity, which is a state of passivity where consumption and enjoyment are delegated \citep{Pfaller2017Interpassivity:Enjoyment}. This passive experience can prevent becoming autonomous because it shifts the emphasis away from evaluating one's abilities and onto the underlying capacity to express desires and needs. To have respect for oneself is to have a perception of oneself as autonomous, which is arguably an essential condition for full autonomy. Respecting oneself entails treating oneself as an equal among others and fully acknowledging the significance of one's abilities. Lack of self-respect can be aggravated by the exacerbation of inferiority complexes, e.g., through unequal access to technology (exacerbating pre-existing ones and reinforcing their systemic nature) and through the way it is utilized (potential to limit exposure to different cultural models and social support for it). In this respect, we can think about this by asking:
    \begin{itemize}
        \item can the system alter the self-perception of the end user? If so, how?
    \end{itemize} 
    
    \item \textbf{Devaluation of autonomy}. ML-based businesses see higher revenues as their usage of ML products grows. In those cases where this materializes into apps, software as a service, or APIs, for example, the quest for engagement is inherent. However, this fundamental dependence on user interaction devaluates the pursuit of autonomy, exemplified by features such as autoplay and endless scrolling, whose nature relies on the default decision to keep the user engaged with the system, ultimately aggravating power asymmetries. The built-in depreciation of autonomy impedes the process of developing competence by lowering the importance ascribed to the development of this critical capacity. This devaluation creates a barrier, preventing individuals from putting enough emphasis on developing their competencies for autonomy and suppressing overall development in the sphere of autonomy. In this context, we can consider this by asking the question:
    \begin{itemize}
        \item does the system make default decisions for the end user? If so, which is the expected impact of those decisions and how do you assess it?
    \end{itemize}
       
    \item \textbf{Respect everybody's autonomy.} People with lower capacity for self-determination (e.g., due to age or medical conditions) have the right to take part in the determination of their own life~\citep{Laitinen2021AIAutonomy}. However, ML implementations could reduce their already limited autonomy by introducing new obstacles.
    Communication, as \cite{Feynman1983FunPushing} pointed out, relies on the premise that the speaker and the recipient have common ground. ML solutions have quickly blended into our daily lives, outpacing social literacy. Those who have not kept up with the times struggle to appraise the content they receive due to a lack of necessary expertise, restricting their autonomy in properly employing such technologies. Besides the initial optimism, the expectations surrounding technological transformation have not materialized~\citep{Muniz2019OrdenCambio}. Instead, over the last two decades, there has been a strong concentration of economic growth and technological innovation in specific geographic hubs, resulting in a partition of the world into two economic spheres. Not all knowledge transcends digital realms; tacit knowledge remains localized, highlighting geography's persistent importance in the technology revolution. Several ML-driven solutions dominate specific domains, with socialization being one of the most notable. In such areas, the options are to either maintain one's autonomy and refrain from engaging in that social arena or to conform and become a member of the dominant cultural model~\citep{Arroyo2022CognitiveSurvival}. Thinking about this aspect, we can address it by posing the question:
    \begin{itemize}
        \item is the tool designed to cater to all age groups and medical conditions for the intended target users? If not, which demographics are left out and why? What is the expected impact?
    \end{itemize}

    \item \textbf{Excessive confidence.} Another important element influencing the state of ML advancement at the moment is the conversation that society is having about ML. Public impressions are shaped in part by the intense excitement surrounding ML. A broadly accepted view presents ML as an impartial reasoning tool that can solve problems that have a universal nature. The appearance of objectivity and universality in ML might encourage over-reliance, prompting individuals to refrain from arguing for their self-determination. The belief that ML can make decisions on their behalf may discourage people from actively participating in the process. This possible over-reliance may inhibit people from exercising control over their choices and actively participating in decision-making processes that shape their lives. In light of this, we could delve into it by asking:
    \begin{itemize}
        \item do you communicate in an easily comprehensible manner the limitations of the system? If so, how have you assessed the comprehensibility of the communication? And if not, why?
    \end{itemize}

\end{itemize}

\subsection{Impact on knowledge representation}
Our goal in this step of the ML pipeline is to model the world by representing it in a way that allows for ML processing. During this step, it is critical to consider how knowledge representation influences individual decision-making processes. The main concerns encountered during this phase are potential misalignments between the captured information and our initial aims, as well as how the heteronomy dynamics may affect the representation of knowledge for the end-user. It is critical to be aware of these potential issues since they might have a major influence on the accuracy and relevance of the knowledge used to inform subsequent decisions within the ML framework. The following possible complications are identified in this phase of ML.

\begin{itemize}
    \item \textbf{Inaccurate data.} Using inaccurate data in ML poses the danger of creating a misalignment between both the desired and the actual reality. The discrepancy, caused by the usage of wrong information, can result in flawed models and predictions, undermining the accuracy of the ML system’s insights and diminishing the well-informedness of end-users. As a result, users who rely on this information for decision-making may be misled or make poor decisions as a result of skewed or incorrect insights obtained from defective data. Inaccurate data can also be generated when information cannot be adequately gathered, either owing to its inaccessibility or privacy limitations, resulting in aggregated values that introduce undesirable noise. Furthermore, data quality may be harmed by missing values or uncertainties, also resulting in inaccurate data. Thinking about this aspect, we can address it by posing the question:
    \begin{itemize}
        \item do you implement data quality controls that account for representativeness and exogenous variables that might influence the results? If so, which ones? If not, why not?
    \end{itemize}
    
    \item \textbf{Space of outcomes.} The choice of knowledge representation that defines the target label in supervised learning has a significant impact on the space of outcomes of ML systems. This is fundamentally finite in classification and recommendation systems and introduces the concern of incompleteness. The technique in which knowledge representation is done may not capture all of the information, resulting in potential gaps. This partial representation can result in a simpler decision-making environment, which influences how end users identify all the options that conform the decision space. The simplification influences the decision-making process by restricting the number and content of the options given to the end user. As a result, the complexities of representing the target label directly contribute to the complexity and richness of the decision space within a certain context. Given this scenario, we can explore it by asking:
    \begin{itemize}
        \item is the system's outcome space rich enough to allow the end user to clearly define the relevant decisions to be made? If so, how was this validated?
    \end{itemize}
    
    \item \textbf{Target audience.} The process of knowledge representation includes determining what information is included and what is excluded for the intended use of the ML solution. The ML system only processes data that is explicitly represented, treating anything else as unknown to the system. As a result, the intended use explicitly and implicitly defines the system's target audience. A lack of clarity around this information may create a potential threat to end-user autonomy, since individuals unfamiliar with the technology may mistakenly obtain access. The lack of explicit parameters for the inclusion or exclusion of specific user groups can impair the system's ability to identify and respond correctly. As a result, end-user autonomy is jeopardized since the system may generate predictions that are completely irrelevant or improper, thus perverting the act of executing decisions. Considering this aspect, we can delve into it by asking:
    \begin{itemize}
        \item do you clearly communicate who the intended audience is for your solution? Do you specify the risks if the system is used for unintended audiences? If so, how?
    \end{itemize}
\end{itemize}

\subsection{Impact on the ML model}
During this stage of the ML pipeline, the modeling logic is applied, building on the knowledge representation chosen. The primary concerns at this stage revolve around the ML solution's ability to impose algorithmic reasoning priming human reasoning. This suggests that the ML model's restrictions may constrain end-user logic. As a result, at this phase, we predominantly address precautions resulting from heteronomy as well as concerns arising from misalignment owing to constraints in knowledge representation. The following factors to watch are identified in this phase of ML.

\begin{itemize}
    \item \textbf{Miss out on significant opportunities.} Transparency is commonly acknowledged as an essential element of reliable ML models. Nonetheless, there is substantial discussion regarding what aspects should be transparent. One critical feature is allowing end users to challenge the suggested outcomes. When this functionality exists, end users may actively alter ML suggestions, ensuring that decisions are aligned with their preferences and beliefs. However, this ideal scenario is still far from being fulfilled. The lack of actionable explanations leads to the possibility that end users will miss out on significant opportunities because they were not integrated into the algorithm design, so lowering the level of comprehension by users. Contemplating this, an approach is to consider asking:
    \begin{itemize}
        \item does the offered solution allow for generating explanations about the reasons behind the prescriptions made in a way that the end user can react? If so, how?
    \end{itemize}
    
    \item \textbf{Opt-out.} An ML system that properly upholds respect for human values should be capable of suggesting alternatives to its use. Specifically, in the European context to be compliant with the General Data Protection Regulation (GDPR) or in the US state of California to be compliant with California Consumer Privacy Act (CCPA), it should provide an opt-out provision. Currently, there is a significant lack of standardized procedures to facilitate such requests, and various practical issues remain unresolved. One such obstacle is the concept of machine unlearning, in which the question of what it means in a logical language for an ML system to forget is still unsolved, thus limiting the decisions one can make. When examining this, we could question:
    \begin{itemize}
        \item does the system integrate effective methods for individuals to be forgotten by the system? If so, which ones?
    \end{itemize}
    
    \item \textbf{Limit access to information.} In ML algorithms, personalization refers to the idea of prioritizing search optimization's exploitation potential over exploration. This tendency is widespread in ML applications since it improves accuracy and other metrics that are frequently used for assessment. However, this strategy fails to uphold fundamental human ideals like autonomy. Prioritizing personalization in situations such as recommendation systems may limit access to information and create echo chambers and filter bubbles that reduce exposure to a variety of viewpoints and information. Furthermore, when it comes to personalization, uncertainty becomes especially important when dealing with individual preferences, which are inherently volatile and changeable. In cases characterized by high uncertainty, algorithms should be able to \textbf{withhold from responding}. Currently, however, algorithms are frequently forced to give a result regardless of the underlying uncertainty, thus affecting the space of outcomes offered to the end-user. Regarding the former point of attention, we might ponder by asking:
    \begin{itemize}
        \item is the model primarily optimized for personalization? If so, which are the potential consequences of limited diversity in the system's outcomes?
    \end{itemize}
    
    \item \textbf{Outsourcing models.} Externalising models is becoming more and more common in ML. It makes it possible to reuse modeling techniques that would be expensive to replicate in-house. This approach comprises using third-party libraries, external APIs, ML as a service, and adopting foundational models. Despite the fact that many of these resources are in the public domain, they frequently lack thorough documentation or validation outlining their limits. As a result, reusing such resources requires a high level of trust. However, the lack of precise knowledge about what is being reused may jeopardize the information available to ML practitioners. They may mistakenly use tools with restrictions they are unaware of, weakening their ability to make educated decisions and choose resources and spreading to the final user this lack of knowledge. Contemplating this, an approach is to consider by asking:
    \begin{itemize}
        \item are all third-party components used in the system identified? If so, which are their limitations?
    \end{itemize}
    
\end{itemize}

\subsection{Impact on the prediction}
At this stage our attention shifts to monitoring the system's usage over time and its impact on decision-making. The principal elements to be aware of arise from how the ML system's choices influence the decision-making processes. These implications fall into three categories: heteronomy, over-reliance, and their impact on the development of autonomy competence. The following factors are identified in this phase of ML.

\begin{itemize}
    \item \textbf{Deskilling.} Overreliance on ML may result in a continued dependence on ML solutions. When people accept ML-generated responses as their own decisions, ML effectively gains decision-making authority. While delegation may be an intentional choice, the focus here is on the consequences of such behavior. Allowing ML to make judgments on someone's behalf suggests that that person may eventually lose the ability to reflect on those decisions. When taken to the extreme, this process might lead to deskilling, in which people gradually lose competency in specific decision domains. This is not necessarily a bad thing, because technology like ML can help you solve problems more efficiently. However, it is critical to identify certain domains where deskilling could cause big problems. Sensitive sectors like health, education, and security will remain dependent on human judgment, at least at some level, to guarantee the development of autonomy competencies. In light of this, one way to approach is by posing the question:
    \begin{itemize}
        \item is the ML solution used in spaces that demand specific technical or scientific knowledge? If so, which are the requirements to ensure that users have the essential competencies to utilize the tool established?
    \end{itemize}
    
    \item \textbf{Informational and decisional privacy.} Informational privacy is synonymous with data privacy, whereas decisional privacy focuses specifically on an individual's ability to make critical decisions without external influence. The way individuals engage with ML systems can differ based on how these systems utilize data. For instance, having control as end-users over whether and to what extent others may comment, interpret, change, or in any other way interfere with how one leads their life may heavily affect the exercise of individuals' autonomy. Thus, having informational and decisional privacy will shape the user experience and interactions with the ML environment. Uncertainty over the level of personal exposure during interactions with a system may cause people to hesitate and decide not to use it. Regardless of one's intention to use the system, fear of negative outcomes may lead to acts that contradict one's preferences, limiting one's ability to act in accordance with one's agency. Given this scenario, we can explore it by asking:
    \begin{itemize}
        \item does the system have the capacity to affect people's informational or decisional privacy? If so, how?
    \end{itemize}
    
    \item \textbf{Performative predictions.} Individuals evolve throughout time and the represented features that formerly correctly reflected them may become obsolete. To effectively support people's self-determination, systems must be adaptable enough to cope with these data shifts. One obstacle emerges from performative predictions, which directly alter and shape what they seek to forecast. This form of prediction has the potential to compromise self-determination by reinforcing a perspective through feedback loops, consequently failing to adjust to users' evolving data and their corresponding varied viewpoints. Thinking about this aspect, we can address it by posing the question:
    \begin{itemize}
        \item do the input data of the system rely on prior modelizations? If so, how are the potential feedback loops handled?
    \end{itemize}
\end{itemize}

The points of focus and questions raised throughout this section help formulate a central inquiry: to what extent is human autonomy vested in the ML solution? Importantly, the end-user should play the leading role in determining decision-making authority.

\section{Conclusions}
\label{sec:conclusions}
The concept of autonomy, which is critical to society in democratic environments due to its critical significance in shaping human relationships and decision-making processes, has received little attention in the field of ML. This weakness is especially notable given ML solutions have a growing impact on different aspects of daily life, ranging from influencing individual decisions to changing larger societal dynamics. A more thorough investigation and integration of autonomy into the ML discourse are critical for supporting responsible and ethically grounded development and deployment of ML systems that are consistent with societal norms and expectations.

The essence of this proposal is its diagnostic nature, which aims to serve as a guideline for those involved in the ML ecosystem. Its major goal is to make it easier to identify potential aspects to monitor concerning autonomy as they occur across the ML pipeline. By using a diagnostic approach, this proposal aims to provide a method of assessing and comprehending the effects of ML technologies on end-user autonomy. Through this diagnostic perspective, the proposal seeks to contribute not only to challenge identification, but also to the growth of informed and ethical decision-making practices in the development and deployment of ML systems.

This approach also seeks to bridge the language gap that exists between individuals from different areas of study. Recognizing that each discipline frequently has its own specialized vocabulary and communication peculiarities, this proposal aims to streamline and unify the language employed, allowing for simpler and more accessible communication across areas borders. The goal is not only to improve communication efficiency, but also to foster interdisciplinary collaboration and contribute to the development of holistic and well-rounded solutions to complex problems spanning multiple fields of expertise.

While there is still work to be done, we hope that this effort contributes to increasing knowledge and literacy about the critical role of autonomy in the ML pipeline. The goal of shining light on this essential ethical principle is to provide ML practitioners with a better grasp of the implications of their solutions for autonomy, and all the posed questions must serve to strengthen the importance of the overarching reflection: what level of autonomy do we want to delegate to AI? Thus, we hope to encourage a more conscientious and ethically grounded approach within the ML community, fostering a collective commitment to responsible innovation and the recognition of autonomy as a critical value to reflect on in the design and deployment of ML solutions. In the end, owing to their inherent authority, the end-user must be the one having effective decision-making authority.

\section{Acknowledgements}
This work is partially supported by MCIN/ AEI/10.13039/501100011033/ under project PID2022-136436NB-I00.

\bibliographystyle{unsrtnat}

\end{document}